\documentclass[twocolumn,showpacs,amssymb,prl,nofootinbib]{revtex4}

\usepackage{graphicx}% Include figure files
\usepackage{dcolumn}% A lign table columns on decimal point
\usepackage{bm}% bold math

\def\gtap{\mathrel{ \rlap{\raise 0.511ex \hbox{$>$}}{\lower 0.511ex
   \hbox{$\sim$}}}} 
\def\ltap{\mathrel{ \rlap{\raise 0.511ex
    \hbox{$<$}}{\lower 0.511ex \hbox{$\sim$}}}} 
\newcommand{\bea}{\begin{eqnarray}} 
\newcommand{\eea}{\end{eqnarray}}
\def\beq{\begin{equation}}
\def\enq{\end{equation}}
\def\ba{\begin{eqnarray}}
\def\ea{\end{eqnarray}}
\def\<{<\!\!}
\def\>{\!\!>}
\def\<{\langle}
\def\>{\rangle}

\def\mdm{\mbox{$m_{\mbox{}_\mathrm{DM}}$}}
\def\gammaann{\mbox{$\Gamma_{\mathrm{ann}} \;$}}

\begin{document}

\vskip-6pt \hfill  \hspace{15truecm}
\begin{tabular}{l}
                 {IPPP/07/34} \\
                 {DCPT/07/68}
                 \end{tabular}

\input{epsf}

\vspace{0.6cm}

\title{Reconstructing WIMP properties with neutrino detectors}

\author{Olga Mena$^1$, Sergio Palomares-Ruiz$^2$ and Silvia
  Pascoli$^2$} 

\affiliation{$^1$ INFN Sez.\ di Roma,
Dipartimento di Fisica, Universit\`{a} di Roma``La Sapienza'', P.le
A.~Moro, 5, I-00185 Roma, Italy} 
\affiliation{$^2$IPPP, Department of Physics, Durham University,
  Durham DH1 3LE, United Kingdom}               

\begin{abstract}
If the dark matter of the Universe is constituted by weakly
interacting massive particles (WIMP), they would accumulate in the
core of astrophysical objects as the Sun and annihilate into particles
of the Standard Model. High-energy neutrinos would be produced in the
annihilations, both directly and via the subsequent decay of leptons,
quarks and bosons. While \v{C}erenkov neutrino detectors/telescopes
can only count the number of neutrinos above some threshold energy, we
study how, by exploiting their energy resolution, large magnetized
iron calorimeter and, possibly, liquid argon and totally active
scintillator detectors, planned for future long baseline neutrino
experiments, have the capability of reconstructing the neutrino
spectrum and might provide information on the dark matter
properties. In particular, for a given value of the WIMP mass, we show
that a future iron calorimeter could break the degeneracy between the
WIMP-proton cross section and the annihilation branching ratios,
present for \v{C}erenkov detectors, and constrain their values with
good accuracy.
\end{abstract}

\maketitle

\section{Introduction} 

Cosmological and astrophysical observations provide growing evidence
of the existence of Dark Matter (DM) in our Universe. One of the
favored candidates is a weakly interacting massive particle (WIMP), a
new stable neutral particle with typical mass in the 10~GeV--TeV range
which interacts weakly~\cite{reviews}. These particles arise naturally
in various extensions of the Standard Model of Particle Interactions
(SM), e.g. the lightest neutralino in supersymmetric models with
R-parity conservation~\cite{reviews}, stable scalars in little Higgs
models~\cite{littleHiggs}, the lightest Kaluza-Klein excitation of the
SM fields in models with universal
extra-dimensions~\cite{extraD}. Direct searches for WIMPs look for the
recoil energy of nuclei due to interactions with the DM particles
passing through the Earth~\cite{directDM}. WIMP signatures can also be
found indirectly, namely by observing gamma-rays, positrons,
anti-protons and neutrinos produced in dark matter
annihilations~\cite{reviews}. As the annihilation rate scales with the
square of the dark matter density, indirect signals are expected to be 
stronger where DM accumulates, as, for instance, the galactic center
and astrophysical bodies. High-energy neutrinos are produced in the
annihilation either directly or by subsequent decay of SM fermions and
bosons~\cite{DMnus} and their spectrum could provide information on
the WIMP mass, the WIMP-proton cross section and the branching ratios
of the various annihilation channels, and hence, on the DM properties.

Searches for high energy neutrinos can be performed in \v{C}erenkov
neutrino detectors/telescopes~\cite{sklimit,nutelescopes}, as
Super-Kamiokande (SK), AMANDA, IceCube, but no positive evidence has
been found so far. However, these detectors are unable to fully
reconstruct the neutrino energy and therefore cannot provide
information on the neutrino spectrum. In the case of small detectors
with large photo-multiplier coverage, like SK, the size limits the
maximum energy for which the events are contained, so only
through-going muons are of interest for energies of tens of GeV or 
higher. In the case of neutrino telescopes like IceCube, least
granular, the size limits the energy threshold for detection. In both
cases and for the  energies of interest, these detectors can only
count neutrinos above their energy threshold. On the other hand,
different conceptual designs for detectors for future neutrino long
baseline experiments show that Magnetized Iron Calorimeter Detectors
(MIND)~\cite{MIND}, or its variants as MONOLITH~\cite{monolith} or
INO~\cite{ino}, Liquid Argon Time Projection Chambers
(LArTPC)~\cite{LAr} and Totally Active Scintillator Detectors
(TASD)~\cite{TASD} are able to measure the energy and direction of the
initial neutrino with good precision at energies of tens of GeVs.

In the present letter, we restrict our analysis to MIND, which are
conceptually similar to the existing MINOS detector~\cite{MINOS} but
with a mass one order of magnitude larger. Due to its ability for a
precise muon charge discrimination, they have been proposed as
ideal detectors for future neutrino factories. For the same reason,
these detectors have also been considered as atmospheric neutrino
observatories, opening up the possibility of determining the neutrino
mass hierarchy and the small mixing angle $\theta_{13}$, due to matter
effects in the oscillations of upward-going atmospheric
neutrinos~\cite{atmos}. Here, we reveal a novel aspect of
MIND~\cite{LArDM}, showing that they might constitute excellent
observatories to detect neutrinos from WIMP annihilations. For a given
value of the WIMP mass, we study the neutrino signal from WIMP
annihilations in these detectors, and exploit their energy and
angular resolution in order to reconstruct the neutrino spectrum and,
ultimately, to study DM properties~\cite{BEK97}. By measuring the
branching ratios of annihilation into different channels and the
WIMP-proton cross section, these measurements could provide a unique
opportunity for getting information on the WIMP couplings with the
particles of the SM.

\section{Neutrinos from WIMPs annihilations} 

When a WIMP goes through a celestial body, as the Sun, it might
interact elastically with the nuclei and get scattered to a velocity
smaller than the escape velocity, remaining gravitationally trapped in
the body. It will then undergo additional scatterings, settling in the
Sun core, giving raise to an isothermal distribution. The WIMPs so
accumulated can annihilate with other WIMPs into SM particles, as
quarks, leptons and, if kinematically allowed, gauge and Higgs
bosons. Typically in the Sun, for sufficiently high capture rate and
annihilation cross section, equilibrium is reached and the
annihilation rate \gammaann is related to the capture rate $C_\odot$
as $\Gamma_{\mathrm{ann}} = 1/2 C_\odot$, which, for the Sun, is given
by~\cite{Gould92}
\ba
C_\odot & \simeq & 9 \times 10^{24} \; {\rm s}^{-1} \,
\left(\frac{\rho_{\rm local}}{0.3 \,  
{\rm GeV/cm}^{3}}\right) \, \left(\frac{270 \, {\rm
    km/s}}{\bar{v}_{\rm local}}\right)^3 \nonumber \\ 
& & \times \left(\frac{\sigma}{10^{-2} \, {\rm pb}}\right)
    \; \left(\frac{50 \, {\rm GeV}}{m_{\rm DM}}\right)^2 ~,
\ea
where $\rho_{\rm local}$ is the local WIMPs density, $\bar{v}_{\rm
local}$ is the velocity dispersion of WIMPs in the halo, $\sigma$ is
the WIMP-proton cross section, and $\mdm~$ is the WIMP mass. The
capture rate can be different by several orders of  magnitude for
spin-dependent and spin-independent interactions. Typically, for most
neutralino models as well as for Kaluza-Klein DM, the spin-dependent
cross-section dominates over the spin-independent one. The opposite
happens for scalar WIMPs. These scattering cross sections are
constrained in direct DM searches~\cite{directDMbounds}, which put
very strong bounds on the spin-independent ones. Since, in the Earth,
the abundance of nuclei with odd mass number is extremely small, the
neutrino flux from WIMP annihilations is strongly constrained by the
bounds on the spin-independent cross section and would be too low to
be interesting for detection in MIND, LArTPC or TASD. Thus, for our
purposes we will focus on WIMP annihilations in the Sun, for which the
much more weakly constrained spin-dependent cross section may play the
dominant role. 

High-energy neutrinos would be produced directly or indirectly via the
decay of other products of the annihilations~\cite{reviews}. A broad
spectrum of neutrinos is so generated and depends on the WIMP mass and
on the branching ratios into the various channels: 
\beq
\frac{dN_\nu}{d\Omega\ dt\ dE_\nu}=\frac{\Gamma_{\mathrm{ann}}}{4\pi
  R^2}\sum_i {\rm BR}_i \frac{dN_{i}}{dE_\nu}~, 
\label{eq:fluxes}
\enq
where the sum includes the possible annihilation channels with
spectrum $dN_i/dE_\nu$ and branching ratio ${\rm BR}_i$, and $R$ is
the Sun-Earth distance. If the DM is a Majorana fermion, the
annihilation amplitude into fermion pairs is proportional to the
fermion mass, so that the neutrino channel is irrelevant and
annihilations are dominated by heavy fermions, $b$s, $\tau$s, $c$s
and, if kinematically allowed, $t$s. Neutralinos can annihilate into
fermions as well as gauge bosons and Higgs and the dominant channels
are controlled by the neutralino composition~\cite{reviews}.
Typically, for neutralinos lighter than the W boson, the annihilation
into $b\bar{b}$ pairs gives the main contribution, with harder
neutrinos from $\tau \bar{\tau}$. For higher masses, the gauge bosons
channels are allowed and can have the highest branching ratio.
Kaluza-Klein WIMPs can annihilate directly into
neutrinos~\cite{extraD} with a branching ratio of few per cent and
would provide a specific signature with a peak in the neutrino
spectrum at $E_\nu=\mdm$. Their main annihilation channels are charged 
leptons and light quarks with the neutrino flux generated dominantly
by the $\tau \bar{\tau}$ mode. In the following, we will treat the
branching ratios as free parameters as their exact values depend on
the type of particle considered and on its couplings.  

Annihilations into $\tau$'s and gauge bosons produce a hard spectrum
which peaks around 0.4~\mdm (0.5~\mdm) for $\nu_{e, \mu}$
($\nu_\tau$) for the former channel and even at higher energies for
the latter~\cite{Cirelli:2005gh}. Quarks hadronize producing a large
number of mesons and baryons which subsequently decay producing a
softer neutrino spectrum. For instance, neutrinos from $b$s and $c$s
have a spectrum which peaks at energies $\sim$~0.15~\mdm~ or lower,
depending on the WIMP mass, and drops to zero around
$E_\nu \sim $~(0.6--0.7)~\mdm~\cite{Cirelli:2005gh}. Light quarks
hadronize mostly in pions which, as muons, get stopped before decaying
and therefore produce only low energy neutrinos. Once produced,
neutrinos propagate in the Sun, being absorbed via charged-current
interactions and loosing energy due to neutral-current ones. These
effects, in addition to neutrino oscillations between different
flavors, need to be taken into account~\cite{Cirelli:2005gh}.

\section{Reconstructing DM properties} 

High-energy neutrinos can be detected in neutrino detectors and
stringent bounds have already been set~\cite{sklimit} and will be
further improved by neutrino telescopes~\cite{nutelescopes}. However,
as these experiments cannot reconstruct the neutrino energy, they can
only provide limited information on the neutrino spectrum. In order to
reconstruct the \emph{neutrino spectrum}, the analysis presented here
exploits the \emph{energy and angular resolution} of MIND, which allow
not only to measure precisely the energy and direction of muons
produced by the charged current interactions (CC) of $\nu_\mu$ and
$\bar{\nu}_\mu$ but possibly also the energy of the
electrons~\cite{MIND,anselmo}, produced by $\nu_e$ ($\bar{\nu}_e$)
interactions. In addition, these detectors will be able to reconstruct
the energy and angle of the hadron shower, so that a good energy and
direction resolution for the incoming neutrino will be achieved. We
show that the degeneracy among different parameters can be broken by
using the muon and electron energy spectra. In particular, a larger
annihilation rate \gammaann could always be traded by a smaller
branching ratio to a hard channel ($\tau^{+} \tau^{-}$) in absence of
information on the spectrum.

In the following we will consider a light WIMP candidate with $\mdm <
80$~GeV and therefore, we will focus on its dominant annihilation
modes for these masses, i.e., $\tau^{+}\tau^{-}$ (hard) and
$b\bar{b}$ (soft) channels. Note that the contribution from $c\bar{c}$
is similar to the one of $b\bar{b}$ but with a lower
normalization~\cite{Cirelli:2005gh}. This technique might not be
competitive with neutrino observatories (AMANDA and Icecube) for
WIMP masses $\gtap 100$~GeV, since the typical sizes of future MIND, 
LArTPC and TASD might not allow a precise measurement of high lepton
energies. However, it should be noted that, even for high WIMP masses,
these detectors could be useful for detecting the low energy tail of
the neutrino spectrum~\cite{CMPPprep}. In our study we assume the mass
of the WIMP to be known and, for definiteness, we take $\mdm = 50$~GeV
and $70$~GeV.

Let us now comment on our assumption of taking a given value for the
WIMP mass. By the time the detectors discussed in the present article
are available, information on the WIMP mass might be already available 
from the LHC (Large Hadron Collider) and, possibly, a future ILC
(International Linear Collider) as well as direct and indirect dark 
matter searches. With the start of LHC in 2008, the neutral WIMP
candidate for the dark matter can be indirectly detected in an event
looking for missing energy and missing transverse momentum. By a
detailed determination of the kinematics of the quarks and leptons,
typically it might be possible to measure the WIMP mass with a 10\%
accuracy~(see, e.g., Ref.~\cite{LHCWIMP}). The prospected error
depends strongly on the model of physics beyond the SM and the values
of the parameters. For example, for  the SPS1a benchmark point, the
LSP mass could be determined to be $\mdm= (96\pm 5)$~GeV at the LHC
and $\mdm= (96\pm 0.05)$~GeV at the ILC~\cite{LHCWIMP}. For less
favorable points in the parameter space, worse accuracies will be
reachable. The value for the mass determined in collider searches
could be used as input in our analysis, assuming that the observed
WIMP corresponds to the dark matter particle.

A model-independent determination of the WIMP dark matter mass could
also be obtained in direct DM searches which look for the nuclear
recoil in an interaction between a WIMP and a nucleus in the
detector~\cite{Lewin:1995rx}. The recoil spectrum is strongly
mass-dependent if $\mdm < 100 \ \mathrm{GeV}^2$. For instance, a
superCDMS-like experiment, with exposures of $3 \times 10^3 \ (3
\times 10^4) \ [3 \times 10^5]$~kg~days and a spin-independent cross
section close to the present bound $\sigma_{\rm SI} = 10^{-7} \
\mathrm{pb}$, would be able to measure a light WIMP mass with an
accuracy of $\sim 25\% \ (15\%) \
[2.5\%]$~\cite{Green:2007rb}. Similar conclusions can be drawn for the
case of spin-dependent cross sections with comparable
statistics. Nevertheless, these measurements are affected by the
uncertainties in the WIMP velocity distribution and in the local WIMP
distribution. A method which is independent of the WIMP density and of
the WIMP-nucleon cross section has been also
proposed~\cite{Shan:2007vn}. By comparing the recoil energy spectrum
in direct searches which use different detector materials, it is found 
that a WIMP mass smaller than 50~GeV can be determined with a
1-$\sigma$ error of 20\% if a total statistics of 500 events is
available~\cite{Shan:2007vn}.

Information on the WIMP mass could also be found from indirect dark
matter searches. Some information on the WIMP mass could also be
obtained from the angular distribution of neutrino-induced muons in 
large water-\v{C}erenkov detectors~\cite{EG95}. On the other hand, as
the WIMP density in the center of galaxies is predicted to be
enhanced, a high rate of WIMP annihilation are expected and produce
sizable fluxes of SM particles, as neutrinos, photons, quarks and
leptons. High energy photons can be produced either directly or via
the hadronic interactions in the interstellar medium. In the first
case, the gamma spectrum presents a very clear signature with a peak
at the WIMP mass, allowing its precise determination. In the case of
gamma-rays from primary hadrons, the spectrum has a sharp energy
cut-off in correspondence with \mdm. It has been estimated that GLAST
could constrain the WIMP mass at the 25\% level, for $\mdm = 100$~GeV,
and future Imaging Atmospheric \v{C}erenkov Telescopes, with
sufficiently small energy threshold, could detect the peak from direct
annihilation into photons~\cite{DMgamma}.

In addition, the energy information in the WIMP neutrino signal in the
detectors considered in the present article could allow, in principle,
to infer directly the WIMP mass by exploiting the cutoff in the
spectra. Since the typical sizes of future MIND, LArTPC and TASD might
not allow to have sufficient statistics at the tail of the spectrum,
this measurement could be very challenging. A detailed analysis needs
to be performed, but it is beyond the scope of the present
study~\cite{CMPPprep}.

We have computed the number of electron and muon neutrino-induced 
CC events for \mdm = 50~and 70~GeV, by using the evolved neutrino
spectra from Ref.~\cite{Cirelli:2005gh}. We have, conservatively,
considered nine energy bins (5 GeV bins from an energy of 5 GeV for
\mdm = 50~GeV and 7~GeV bins from an energy of 7~GeV for \mdm =
70~GeV), where the low energy cut is adopted to avoid the atmospheric 
neutrino background. With good statistics, the precision in the
measurement of \mdm, for $\mdm \gtap 50$~GeV, depends on the muon
energy resolution of MIND at energies $E_\nu\sim50-100$~GeV, expected
to be $\sim 5-8\%$ at $50$~GeV and slightly worse for higher
energies~\cite{anselmo}. 

For an ideal detector with perfect lepton angular resolution, the
atmospheric neutrino background would be reduced to the fraction of
events within the angular size of the Sun ($\sim 6.7\times
10^{-5}$~sr). However, the lepton does not travel in the same
direction as the incident neutrino except when averaged over many
events. We conservatively take the root mean square (rms) spread in
direction between the incident neutrino and the lepton (in radians) as 
\beq
\theta_{\rm rms}\simeq \sqrt{\frac{1\ \textrm{GeV}}{E_\nu}}~,
\label{eq:rms}
\enq
where $E_\nu$ is the energy of the incoming neutrino. Therefore, a
very conservative approach is to sum over the atmospheric neutrino
background contributions within a region of opening angle
$\theta_{\rm rms}$.

The expected number of muon (electron) neutrino-induced events in the
$i$-th bin is computed as a function of the WIMP-proton scattering
cross section and of the branching ratios into $b^{+}b^{-}$
(soft channel) and $\tau^{+}\tau^{-}$ (hard channel),   
\ba
N_{i,\mu(e)} & = & N_{\rm T} \, t \, \int_{E_i}^{E_i
  +\Delta}dE_\nu \int d\theta \, \left(\phi_{\nu_\mu(\nu_e)} (E_\nu) \,
\sigma^{\rm CC}_{\nu_\mu(\nu_e)} (E_\nu) \right. \nonumber \\
 & & \left. + \, \phi_{\bar{\nu}_\mu(\bar{\nu}_e)} (E_\nu) \,
\sigma^{\rm CC}_{\bar{\nu}_\mu(\bar{\nu}_e)} (E_\nu) \right) \times
\frac{V_{\mu(e)}}{V_{\rm det}}~,
\label{eq:events}
\ea
where $\Delta$ is the energy bin width, $\theta$ is the angle which
measures the relative position of the Sun with respect to the detector
location, $N_{\rm T}$ is the number of available targets, $V_{\rm
  det}$ is the total volume of the detector, $t$ is the exposure time,
$\phi$ is the evolved (anti)neutrino spectra, $\sigma^{\rm CC}$ is the
CC (anti)neutrino cross section and $V_{\mu, e}$  is the volume of
detector available for the neutrino to interact. For electron-like
events, which are all contained, this fiducial volume $V_e$ is simply
the detector volume $V_{\rm det}$. For muon-like contained-events, it
depends on the detector geometry and on the muon range in iron $R_\mu
(E_\mu)$. For a detector with cylindrical shape, it is given
by~\cite{AS01}    
\ba
V_\mu(E_\mu,\theta) & = & 2 h r^2
\arcsin\left(\sqrt{1-\frac{R^2_\mu(E_\mu)}{4 \, r^2}\sin^2 \theta}
\right) \nonumber \\
& & \times \left(1-\frac{R_\mu(E_\mu)}{h}|\cos\theta|\right)~, 
\label{eq:range}
\ea
where $r=13$~m and $h=20$~m are the detector radius and height we have
considered. We approximate the muon energy by its averaged value,
which for the energy range of interest is $\langle E_\mu \rangle =
0.48 \, E_\nu$~\cite{GQRS95}. Also notice that the volume of detector
available is a function of the azimuthal angle $\theta$, which varies
from 0 to $\pi$ in 12 hours. Thus, for the assumed geometry, the
effective volume will be maximal when the Sun is at the horizon.  On
the other hand, if a good energy resolution is achieved up to some
energies even for partially-contained events (by means of using the
bending of the muon trajectory due to the magnetic field), the
fiducial volume to be considered is $V_{\rm det}$. We will consider
both cases, when partially-contained muon-like events are not included
in the analysis and when all muon-like events with the interaction
vertex in the detector contribute.

The expected number of muon and electron neutrino-induced events is
then fitted performing a $\chi^2$ analysis. We define
\beq
\chi_{\mu^{\pm} + e^{\pm}}^2 = \sum_{\alpha, \alpha'= \mu,
  e}\sum_{i,j} (n_{i,\alpha} - N_{i, \alpha})
C_{i,\alpha:,j,\alpha'}^{-1} (n_{j,\alpha'} - N_{j,\alpha'})\,, 
\enq
where $C$ is the covariance matrix and only statistical errors, being
the dominant ones, have been considered. $N_{i,\alpha}$ is given
by Eq.~(\ref{eq:events}) and $n_{i,\alpha}$ represents the simulated
muon (electron) data taking into account the total muon or electron
atmospheric neutrino background within the $i$-th energy bin and
integrated over a solid angle $\Delta \Omega= 2\pi \theta_{\rm rms}$,
and is given by
\beq
n_{i,\alpha}=\textrm{Smear}(N_{i,\alpha} + N_{i,{\rm atmos}})
-N_{i,{\rm atmos}}~,  
\enq
where the function Smear indicates that Gaussian or Poisson smearing
has been applied following the Monte Carlo techniques of
Ref.~\cite{PDG} to mimic the statistical uncertainty.

\begin{figure}[t]
\begin{center}
\includegraphics[width=3.5in]{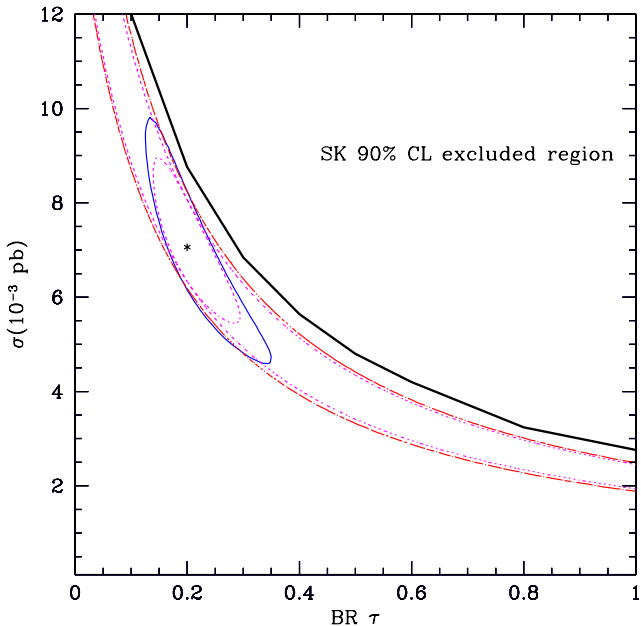}
\caption{\footnotesize 
The solid blue (dotted-dashed red) contours denote the $90\%$ CL
limits, for 2 degrees of freedom, for the simultaneous extraction of
the WIMP-proton cross section and the annihilation branching ratio
into the $\tau^{+} \tau^{-}$ channel with (without) energy
information by using only fully-contained events with an exposure of
750~kton $\cdot$ yrs (see text for details). The simulated nature
values are $\mdm=50$~GeV, $\sigma = 7 \times 10^{-3}$~pb and ${\rm
  BR}_{\tau^+\tau^-}=20\%$. The dashed magenta lines (inner contours)
illustrate the case (for 90\%~CL) in which the atmospheric neutrino
background is integrated only over the real size of the Sun. We also
show the 90\%~CL excluded region by SK~\cite{sklimit}.
} 
\label{fig:fig1}
\vspace{-2mm}
\end{center}
\end{figure}

\begin{figure}[t]
\begin{center}
\includegraphics[width=3.5in]{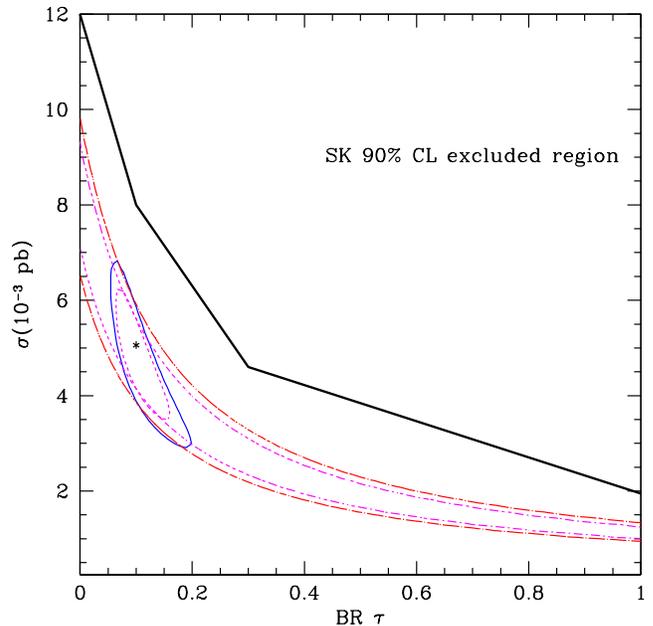}
\caption{\footnotesize The solid blue (dotted-dashed red) contours
  denote the $90\%$ CL limits, for 2 degrees of freedom, for the
  simultaneous extraction of the WIMP-proton cross section and the
  annihilation branching ratio into the $\tau^{+} \tau^{-}$ channel
  with (without) energy information by using only fully-contained
  events with an exposure of 750~kton $\cdot$ yrs (see text for
  details). The simulated nature values are $\mdm =70$~GeV, $\sigma =
  5 \times 10^{-3}$~pb and ${\rm BR}_{\tau^+\tau^-}=10\%$. The dashed
  magenta lines (inner contours) illustrate the case (for 90\%~CL) in
  which partially-contained events are also included in the
  analysis. We also show the 90\%~CL excluded region by
  SK~\cite{sklimit}.
} 
\label{fig:fig2}
\end{center}
\end{figure}

In Figs.~\ref{fig:fig1} and \ref{fig:fig2}, we illustrate the
simultaneous extraction of the scattering cross section of the WIMP
off a proton and the branching ratio into $\tau^{+} \tau^{-}$ assuming
a $50$~kton MIND with the geometry described above. In
Fig.~\ref{fig:fig1} we consider the case of a WIMP mass of \mdm =
50~GeV and the input simulated values are $\sigma = 7 \times 10^{-3}$
pb and BR$_{\tau^+\tau^-}$ = $20\%$. We have assumed 15 years of data
taking and have only included electron-like and contained muon-like
events. The dotted-dashed red lines depict the 90\%~confidence level
(CL) contours assuming that no energy information is available: as
previously anticipated, a larger annihilation rate could always be
mimicked by a lower branching ratio into $\tau^{+} \tau^{-}$. Notice
that there will be a continuous region of degenerate solutions,
implying the impossibility of extracting either the different
branching ratios or the WIMP-proton scattering cross section. In this
scenario, the measurement of the WIMP mass is of course out of
reach. Instead, if we now include the information on the neutrino
energy spectrum, we obtain the $90\%$~CL contours depicted by the
solid blue line. The strong correlation between $\sigma$ and
BR$_{\tau^+\tau^-}$ (or BR$_{b^+b^-}$) is broken. The reason is
simple: the differential neutrino spectrum for each possible
annihilation channel (hard or soft) has a characteristic shape,
different from one channel to another. For this case, \mdm = 50~GeV,
the spectrum for $b\bar{b}$ peaks at $\sim 7$~GeV and represents
roughly $20\%$ of that of $\tau^{+}\tau^{-}$, peaked at $\sim
20$~GeV. We also illustrate the results for the ideal case for which
the atmospheric neutrino background only contributes within the real
angular size of the Sun (dashed magenta lines). As can be seen, by 
considering a more detailed and realistic expression for the angular
resolution than that given by Eq.~(\ref{eq:rms}), our results would
not be significantly improved. In Fig.~\ref{fig:fig2} we show the
90\%~CL contours for 15 years for $\mdm=70$~GeV with the input
simulated values $\sigma= 5 \times 10^{-3}$~pb and BR$_{\tau^+\tau^-}
= 10\%$. The solid blue line represents the case when
partially-contained muon-like events are not considered, whereas the
inner dotted magenta line depicts the case when we also include this
type of events to perform the analysis. In this figure, we have
considered the atmospheric neutrino background integrated over a
half-cone aperture given by Eq.~(\ref{eq:rms}). It is important to
note that by adding the partially-contained events, in order to
achieve the same results as for the case of only contained events,
the total number of years could be reduced from 15 to 10 years of data
taking, the minimum running time of these detectors in the context of
neutrino factories. In both figures, the equivalent $90\%$ CL SK
limit~\cite{sklimit} is included.

As both figures show, for these relatively low WIMP masses and large
spin-dependent cross sections, the determination of the annihilation
branching ratios and WIMP-proton cross section could be
achieved. Notice that a positive signal of neutrinos from WIMP
annihilations in the Sun in these detectors will immediately point to
a spin-dependent cross section. In LArTPC and TASD similar results
could be obtained. In fact, we have checked that considering only the
$\nu_\mu$ or the $\nu_e$ signal does not modify the results
substantially, a part from a small change due to the smaller
atmospheric neutrino background for $\nu_e$. A detailed analysis for
different WIMP masses and type of detectors is in
preparation~\cite{CMPPprep}.

\section{Conclusions} 

Searches of high-energy neutrinos from WIMP annihilations in the Sun
could constitute a powerful probe of WIMP properties. While the total
neutrino flux is controlled by the annihilation rate (proportional to
the WIMP-proton cross section), the shape of the neutrino spectrum
depends on the WIMP mass and on the branching ratios of different
channels. \v{C}erenkov neutrino detectors/telescopes are counting
experiments and can only provide limited information. On the contrary,
we have shown that, by exploiting the good energy and angular
resolution at energies of tens of GeV, future MIND, LArTPC and TASD
could have the capability to reconstruct the neutrino spectrum, and
could provide important information on the WIMP mass, its annihilation
branching ratios and WIMP-proton cross section. In particular, our
analysis shows that, for a given value of the WIMP mass, the
degeneracy between the WIMP-proton cross section and the branching
ratios into soft and hard channels could be broken and important
information on the WIMP dark matter couplings might be obtained. This
information should be combined with the constraints and measurements
from direct and indirect DM searches to probe the nature and
properties of DM particles and to test that the WIMP candidate found
in collider searches does indeed constitute the observed dark matter
of the Universe.

\section*{Acknowledgments} 

It is pleasure to thank A.~Cervera, P.~Lipari and G.~Weiglein for
useful discussions. OM is supported by the European Programme ``The
Quest for Unification'', contract MRTN-CT-2004-503369.  SPR is
partially supported by the Spanish Grant FPA2005-01678 of the MCT. SP
is partially supported by CARE, contract number
RII3-CT-2003-506395. OM and SP would like to thank the Theoretical
Physics Department at Fermilab for hospitality.

\end{document}